\documentclass[final,5p,times]{elsarticle}
\biboptions{sort&compress,numbers}

\usepackage{graphicx}% Include figure files
\usepackage{bm}% bold math
\begin{document}

\begin{frontmatter}

\title{Determination of $\pi\pi$ scattering lengths from measurement 
of $\pi^+\pi^-$ atom lifetime}

\author[s]{B.Adeva} 
\author[d]{L.Afanasyev} 
\author[f]{M.Benayoun} 
\author[zu]{A.Benelli} 
\author[cz]{Z.Berka} 
\author[p]{V.Brekhovskikh} 
\author[b]{G.Caragheorgheopol} 
\author[cz]{T.Cechak} 
\author[jt]{M.Chiba} 
\author[p]{P.V.Chliapnikov} 
\author[b]{C.Ciocarlan} 
\author[b]{S.Constantinescu} 
\author[ba]{S.Costantini} 
\author[b]{C.Curceanu (Petrascu)} 
\author[cz]{P.Doskarova}
\author[itt]{D.Dreossi} 
\author[c]{D.Drijard\corref{cor1}} 
\ead{Daniel.Drijard@cern.ch} 
\author[d]{A.Dudarev}
\author[c]{M.Ferro-Luzzi} 
\author[s]{J.L.Fungueiri\~no Pazos}
\author[c,s]{M.Gallas Torreira} 
\author[cz]{J.Gerndt} 
\author[if]{P.Gianotti} 
\author[ba]{D.Goldin}
\author[s]{F.Gomez} 
\author[p]{A.Gorin} 
\author[d]{O.Gorchakov} 
\author[if]{C.Guaraldo} 
\author[b]{M.Gugiu} 
\author[c]{M.Hansroul} 
\author[czr]{Z.Hons} 
\author[cz]{R.Hosek} 
\author[if,b]{M.Iliescu} 
\author[d]{V.Karpukhin} 
\author[cz]{J.Kluson} 
\author[k]{M.Kobayashi} 
\author[g]{P.Kokkas} 
\author[d]{V.Komarov} 
\author[d]{V.Kruglov} 
\author[d]{L.Kruglova} 
\author[d]{A.Kulikov} 
\author[d]{A.Kuptsov} 
\author[d]{K.I.Kuroda} 
\author[im]{A.Lamberto} 
\author[c,u]{A.Lanaro} 
\author[p]{V.Lapshin} 
\author[cza]{R.Lednicky} 
\author[f]{P.Leruste} 
\author[if]{P.Levi Sandri} 
\author[s]{A.Lopez Aguera} 
\author[if]{V.Lucherini} 
\author[jk]{T.Maki} 
\author[p]{I.Manuilov} 
%\author[sc]{J.Marin\fnref{fn1}}
\author[sc]{J.Marin}
\author[f]{J.L.Narjoux} 
\author[d,c]{L.Nemenov} 
\author[d]{M.Nikitin} 
\author[s]{T.Nunez Pardo} 
\author[jks]{K.Okada} 
\author[d]{V.Olchevskii} 
\author[s]{A.Pazos} 
\author[b]{M.Pentia} 
\author[it]{A.Penzo} 
\author[c]{J.M.Perreau} 
\author[s]{M.Plo} 
\author[b]{T.Ponta} 
\author[im]{G.F.Rappazzo} 
\author[p]{A.Riazantsev} 
\author[s]{J.M.Rodriguez} 
\author[s]{A.Rodriguez Fernandez} 
\author[if]{A.Romero Vidal}
\author[p]{V.M.Ronjin} 
\author[p]{V.Rykalin} 
\author[s]{J.Saborido} 
\author[s]{C.Santamarina} 
\author[be]{J.Schacher} 
\author[ba]{C.Schuetz} 
\author[p]{A.Sidorov} 
\author[cz]{J.Smolik} 
\author[jks]{F.Takeutchi} 
\author[d]{A.Tarasov} 
\author[ba]{L.Tauscher}
\author[s]{M.J.Tobar} 
\author[cz]{T.Trojek} 
\author[m]{S.Trusov} 
\author[d]{V.Utkin}
\author[s]{O.V\'azquez Doce} 
\author[ba]{S.Vlachos} 
\author[d]{O.Voskresenskaya}
\author[cz]{T.Vrba}
%\author[sc]{C.Willmott\fnref{fn1}}
\author[sc]{C.Willmott}
\author[m]{V.Yazkov} 
\author[k]{Y.Yoshimura} 
\author[d]{M.Zhabitsky} 
\author[d]{P.Zrelov} 

\address[s]{Santiago de Compostela University, Spain}
\address[d]{JINR Dubna, Russia}
\address[f]{LPNHE des Universites Paris VI/VII, IN2P3-CNRS, France}
\address[zu]{Zurich University, Switzerland}
\address[cz]{Czech Technical University in Prague, Prague, Czech Republic}
\address[p]{IHEP Protvino, Russia}
\address[b]{IFIN-HH, National Institute for Physics and Nuclear Engineering, Bucharest, Romania}
\address[jt]{Tokyo Metropolitan University, Japan}
\address[ba]{Basel University, Switzerland}
\address[if]{INFN, Laboratori Nazionali di Frascati, Frascati, Italy}
\address[itt]{INFN, Sezione di Trieste and Trieste University, Trieste, Italy}
\address[c]{CERN, Geneva, Switzerland}
\address[czr]{Nuclear Physics Institute ASCR, Rez, Czech Republic}
\address[k]{KEK, Tsukuba, Japan}
\address[g]{Ioannina University, Ioannina, Greece}
\address[im]{INFN, Sezione di Trieste and Messina University, Messina, Italy}
\address[u]{University of Wisconsin, Madison, USA} 
\address[cza]{Institute of Physics ASCR, Prague, Czech Republic}
\address[jk]{UOEH-Kyushu, Japan}
%\address[sc]{CIEMAT, Madrid, Spain\fnref{fn1}} % no mark produced
\address[sc]{CIEMAT, Madrid, Spain${}^1$}
\address[jks]{Kyoto Sangyo University, Kyoto, Japan}
\address[it]{INFN, Sezione di Trieste, Trieste, Italy} 
\address[be]{Bern University, Switzerland}
\address[m]{Skobeltsin Institute for Nuclear Physics of Moscow State University, Moscow, Russia}

\cortext[cor1]{Corresponding author} 

\fntext[fn1]{Associated with the university of Santiago de Compostela
  for technical support in the GEM/MSGC detector}

%\collaboration{DIRAC Collaboration}

\begin{abstract}
  The DIRAC experiment at CERN has achieved a sizeable production of
  $\pi^+\pi^-$ atoms and has significantly improved the precision on
  its lifetime determination. From a sample of
%about 21200
  21227 atomic pairs, a $4\%$ measurement of the S-wave $\pi\pi$
  scattering length difference $|a_0-a_2| =
\left(\left.0.2533^{+0.0080}_{-0.0078}\right|_\mathrm{stat}
\left.{}^{+0.0078}_{-0.0073}\right|_\mathrm{syst}\right)
M_{\pi^+}^{-1}$ has been attained,
  providing an important test of Chiral Perturbation Theory.
\end{abstract}

\begin{keyword}

DIRAC experiment
\sep
elementary atom
\sep
pionium atom
\sep
pion scattering

\end{keyword}

\journal{Physics Letters B}

\end{frontmatter}

%\linenumbers

\section{Introduction}

Pionium ($A_{2\pi}$) is the $\pi^+\pi^-$ hydrogen-like atom, with
$378~fm$ Bohr radius, which decays predominantly into
$\pi^0\pi^0$~\cite{Uretsky61}. The alternative $\gamma \gamma$ decay
accounts for only $\sim 0.4\%$ of the total rate~\cite{Gasser08}. Its
ground-state lifetime is governed by the $\pi\pi$ S-wave scattering
lengths $a_I$, with total isospin $I=0,2$~\cite{Uretsky61, Gasser01}:
\begin{equation}
\label{eq:gasser}
\Gamma_{2\pi^0}=
\frac{2}{9} \; \alpha^3 \; p^{\star}(a_0-a_2)^2 (1+\delta)M_{\pi^+}^2,
\end{equation}
where $p^{\star} =
\sqrt{M_{\pi^+}^2-M_{\pi^0}^2-(1/4)\alpha^2M_{\pi^+}^2}$ is the
$\pi^0$ momentum in the atom rest frame, $\alpha$ is the
fine-structure constant, and $\delta=(5.8\pm1.2)\cdot10^{-2}$ is a
correction of order $\alpha$ due to QED and QCD \cite{Gasser01} which
ensures a 1\% accuracy of equation (\ref{eq:gasser}).  The value of
$a_0$ and $a_2$ can be rigorously calculated in Chiral Perturbation
Theory (ChPT)~\cite{Wein79, Gasser85}, predicting $a_0-a_2 = (0.265\pm
0.004)M_{\pi^+}^{-1}$ and the $A_{2\pi}$ lifetime $\tau = (2.9\pm 0.1)
\cdot 10^{-15}$ s~\cite{Colan01NP}.  The measurement of
$\Gamma_{2\pi^0}$ provides an important test of the theory since
$a_0-a_2$ is sensitive to the quark condensate defining the
spontaneous chiral symmetry breaking in QCD~\cite{Knecht95}. The
method reported in this article implies observation of the pionium
state through its ionization into two
%charged 
pions. Given its large Bohr radius, this is directly sensitive to
$\pi\pi$ scattering at threshold, $M_{\pi\pi} \sim 2M_{\pi^+}$, and
thus delivers a precision test of the theory without requiring
threshold extrapolation, as for semileptonic $Ke4$
decays~\cite{Bat10}, or substantial theoretical input as for $K
\rightarrow 3\pi$ decays~\cite{Bat09}.

\section{Pionium formation and decay}
In collisions with target nuclei, protons can produce pairs of
oppositely charged pions. Final-state Coulomb interaction leads to an
enhancement of $\pi^+\pi^-$ pairs at low relative c.m. momentum ($Q$)
and to the formation of $A_{2\pi}$ bound states or pionium. 
%
%% These atoms may either decay mainly into $\pi^{0}\pi^{0}$, or
%% evolve by excitation (de-excitation) to different quantum states and
%% finally decay or be broken up (be ionized) by the electric field of
%% the target atoms. 
%
These atoms may either directly decay, or evolve by excitation
(de-excitation) to different quantum states. They would finally decay
or be broken up (be ionized) by the electric field of the target
atoms.  In the case of decay, the most probable channel is
$\pi^{0}\pi^{0}$ and the next channel is $\gamma\gamma$ with a small
branching ratio of 0.36\%.  In the case of breakup, characteristic
atomic pion pairs emerge \cite{Nem85}. These have a very low $Q$
($<3$~MeV/$c$) and very small opening angle in the laboratory frame
($<3$~mrad). A high-resolution magnetic spectrometer ($\Delta
p/p\sim~3 \cdot 10^{-3}$) is used \cite{Adeva03} to split the pairs
and measure their relative momentum with sufficient precision to
detect the pionium signal. This signal lays above a continuum
background from free (unbound) Coulomb pairs produced in
semi-inclusive proton-nucleus interactions. Other background sources
are non-Coulomb pairs where one or both pions originate from a
long-lived source ($\eta, \eta', \Lambda, \dots$) and accidental
coincidences from different proton-nucleus interactions.

The first observation of $A_{2\pi}$ was performed in the early
1990s~\cite{Afan94}. Later, the DIRAC experiment at CERN was able to
produce and detect $\sim6000$ atomic pairs and perform a first
measurement of the pionium lifetime \cite{Adeva05}.  We now present
final results from the analysis of~$\sim1.5 \cdot 10^9$ events
recorded from 2001 to 2003.  Compared to the results in
\cite{Adeva05}, this analysis has reduced systematic errors and
improved track reconstruction, mostly due to the use of the GEM-MSGC
detector~\cite{Adeva03} information, which leads to a larger signal
yield.  The present data come from collisions of 20 and 24 GeV/$c$
protons, delivered by the CERN PS, impinging on a thin Ni target foil
of 94 or 98 $\mu$m thickness for different run periods.

\section{Pionium detection and signal analysis}
Low relative-momentum prompt and accidental $\pi^+\pi^-$ pairs are
produced at the target and selected by the multi-level trigger when
their time difference, recorded by the two spectrometer arms, is
$|\Delta t|<30$ ns.  A suitable choice of the target material and
thickness provides the appropriate balance between the $A_{2\pi}$
breakup and annihilation yields, with reduced
multiple-scattering~\cite{Afan96, Santa03}.  For a thin Ni target, of
order $\sim 10^{-3}~X_0$, the relative c.m. momentum $Q$ of the atomic
pairs is less than $\sim3$~MeV/$c$ and their number is $\sim 10\%$ of
the total number of free pairs in the same $Q$ region~\cite{Gorch96}.
The experiment is thus designed for maximal signal sensitivity in a
very reduced region of the $\pi^+\pi^-$ phase space. This is done by
selective triggering and by exploiting the high resolution of the
spectrometer and background rejection capabilities.  The longitudinal
($Q_L$) and transverse ($Q_T$) components of $\vec{Q}$, defined with
respect to the direction of the total laboratory momentum of the pair,
are measured with precisions $0.55$~MeV/$c$ and $0.10$~MeV/$c$,
respectively.

The double differential spectrum of prompt $\pi^+\pi^-$ pairs
$N_{\mathrm{pr}}$ (defined by $|\Delta t|<0.5$ ns), composed of atomic
$n_{\mathrm{A}}$, Coulomb $N_{\mathrm{C}}$, non-Coulomb
$N_{\mathrm{nC}}$, and accidental $N_{\mathrm{acc}}$ pairs, can be
$\chi^2$-analysed in the ($Q_T,~Q_L$) plane by minimizing the
expression
\begin{equation}\label{chi2}
\chi^2 = \sum_{ij} \frac 
{[M^{ij} - F_A^{ij}  - F_B^{ij}]^2}
{[M^{ij} + (\sigma_A^{ij})^2 + (\sigma_B^{ij})^2]} ~~.
\end{equation} 
Here
\begin{equation}\label{meq}
M (Q_T,Q_L) = \left(\frac{d^2N_{\mathrm{pr}}}{dQ_TdQ_L}\right) 
\Delta Q_T \Delta Q_L, 
\end{equation}
and the sum in (\ref{chi2}) runs over a two-dimensional grid of
$|Q_L|<15$~MeV/$c$ and $|Q_T|<5$~MeV/$c$, with bin centres located at
values ($Q_T^i,~Q_L^j$) and uniform bin size $\Delta Q_T=\Delta
Q_L=0.5$~MeV/$c$. The $F_A$ and $F_B$ functions describe the $A_{2\pi}$
signal and the $N_{\mathrm{C}}+N_{\mathrm{nC}}+N_{\mathrm{acc}}$
three-fold background, respectively; $\sigma_A$ and $\sigma_B$ are
their statistical errors.  The analysis is based on the
parametrization of $F_A$ and $F_B$ and the precise Monte Carlo
simulation of the detector response.
    
The $F_A$ signal has been simulated ~\cite{Zhab07, Zhab08} according
to an accurate model of $A_{2\pi}$ production,
propagation~\cite{Afan96}, and interaction with the target
medium~\cite{Santa03, Afan99, Afan02, Schum02}.

In the background $F_B$, the $N_{\mathrm{nC}}$ and the
$N_{\mathrm{acc}}$ double differential spectra were parametrized
according to two-body phase space and Lorentz boosted to the
laboratory frame using the observed pion pair spectra ~\cite{Zhab07}.
The spectrum of $N_{\mathrm{C}}$ pairs is enhanced at low $\vec{Q}$
with $Q$ defined at the point of production, by the Coulomb
interaction according to the Gamow--Sommerfeld factor
\begin{equation}\label{sommerfeld}
A_{\mathrm{C}}(Q) = \frac{2\pi M_\pi \alpha/Q}
{1-\exp\left( -2\pi M_\pi \alpha/Q\right) } .
\end{equation}
The finite size of the production source and final-state interaction
effects have been calculated \cite{Led08, Chl09} and applied to
simulated atomic and Coulomb pairs. An additional momentum-dependent
correction has been applied to the simulated $N_{\mathrm{C}}$ spectrum
to take into account a small ($<0.5\%$) contamination, measured by
time-of-flight~\cite{note0702}, due to misidentified $K^+K^-$ pairs.
Small admixtures of misidentified $p\bar{p}$ and residual
contamination from $e^+e^-$ pairs have been measured and produce no
effect on the final result.

\begin{figure}[htb]
\includegraphics[viewport=12mm 35mm 185mm 247mm,width=\columnwidth]{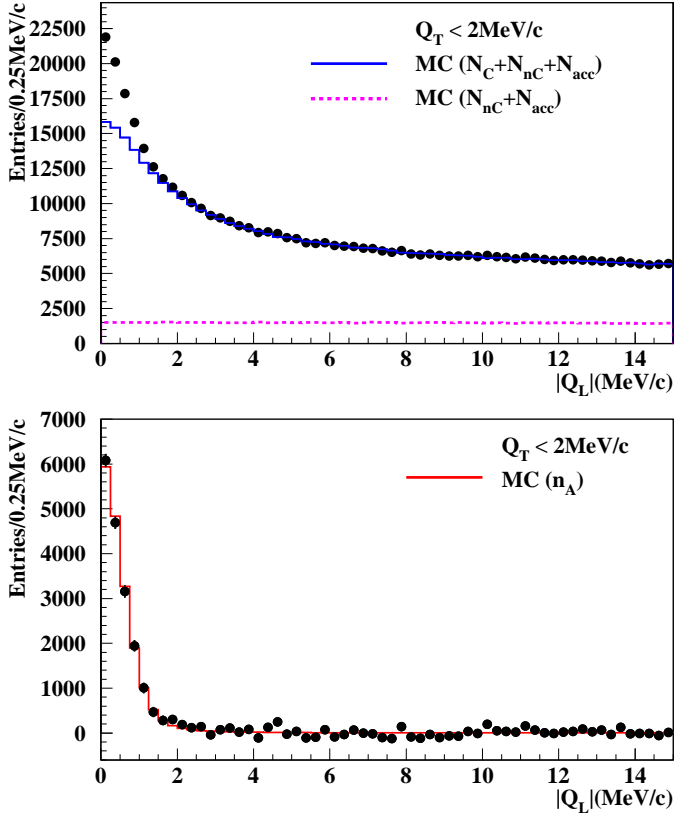}
\caption{$|Q_L|$ fit projections of the $\pi^+\pi^-$ spectrum from 
  data (dots) and simulation (MC lines). The top plot shows the
  experimental spectrum compared with the simulated background
  components (no pionium signal), with (solid line) and without
  (dotted line) Coulomb pairs ($N_{\mathrm{C}}$). The bottom plot
  shows the experimental $|Q_L|$ spectrum after background subtraction
  and the simulated pionium spectrum.}
\label{qldata}
\end{figure}

The fraction of accidental pairs in $F_B$ was measured by
time-of-flight to be $\omega_{\mathrm{acc}}~\simeq12.5\%$, averaged
over the pair momentum and the different data sets.

The experimental resolutions on the momentum and opening angle must be
accurately simulated in order to extract the narrow pionium signal.
Multiple-scattering in the target and the spectrometer is the primary
source of uncertainty on the $Q_T$ measurement. In order to achieve
the desired $Q_T$ resolution, the scattering angle must be known with
$\sim 1\%$ precision, which is beyond the currently available GEANT
description~\cite{Lynch91}.
\begin{figure}
\includegraphics[viewport=11mm 69mm 185mm 211mm,width=\columnwidth]{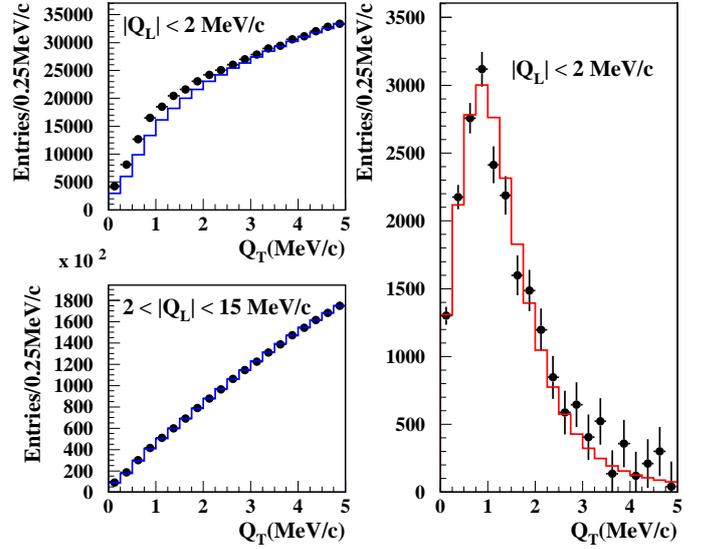}
\caption{$Q_T$ fit projections of the $\pi^+\pi^-$ spectrum from data (dots)
  and simulation (line). The left plots show the comparison between
  the experimental spectra and the full simulated background. The
  plots correspond to different $Q_L$ regions: top left plot in the
  $A_{2\pi}$ signal region (low $|Q_L|$) and bottom left plot away
  from it (higher $|Q_L|$). The right plot shows the $Q_T$ spectrum
  after background subtraction and the simulated pionium spectrum.}

\label{qtdata}
\end{figure}
An improved multiple-scattering description was implemented based on
dedicated measurements of the average scattering angle off material
samples~\cite{note0806}. A cross-check with the standard GEANT
description was made by comparing the momentum evolution of the
measured distance between $\pi^+$ and $\pi^-$ at the
target~\cite{Note-05-16i}.

The $Q_L$ resolution was checked using $\Lambda$ decays with small
opening angle. The widths of reconstructed real and simulated $\Lambda
\rightarrow p\pi^-$ were compared. A $3.4\%$ relative difference was
observed and attributed to residual fringing magnetic field effects,
multiple scattering in the downstream vacuum channel exit window, and
to a small misalignment between the spectrometer arms. Such effects
have been altogether absorbed into an additional Gaussian smearing
term, of width $0.66 \cdot 10^{-3}$, convoluted with the simulated
momentum resolution function.
\begin{figure}
\includegraphics[viewport=11mm 26mm 187mm 295mm,angle=90,width=\columnwidth]{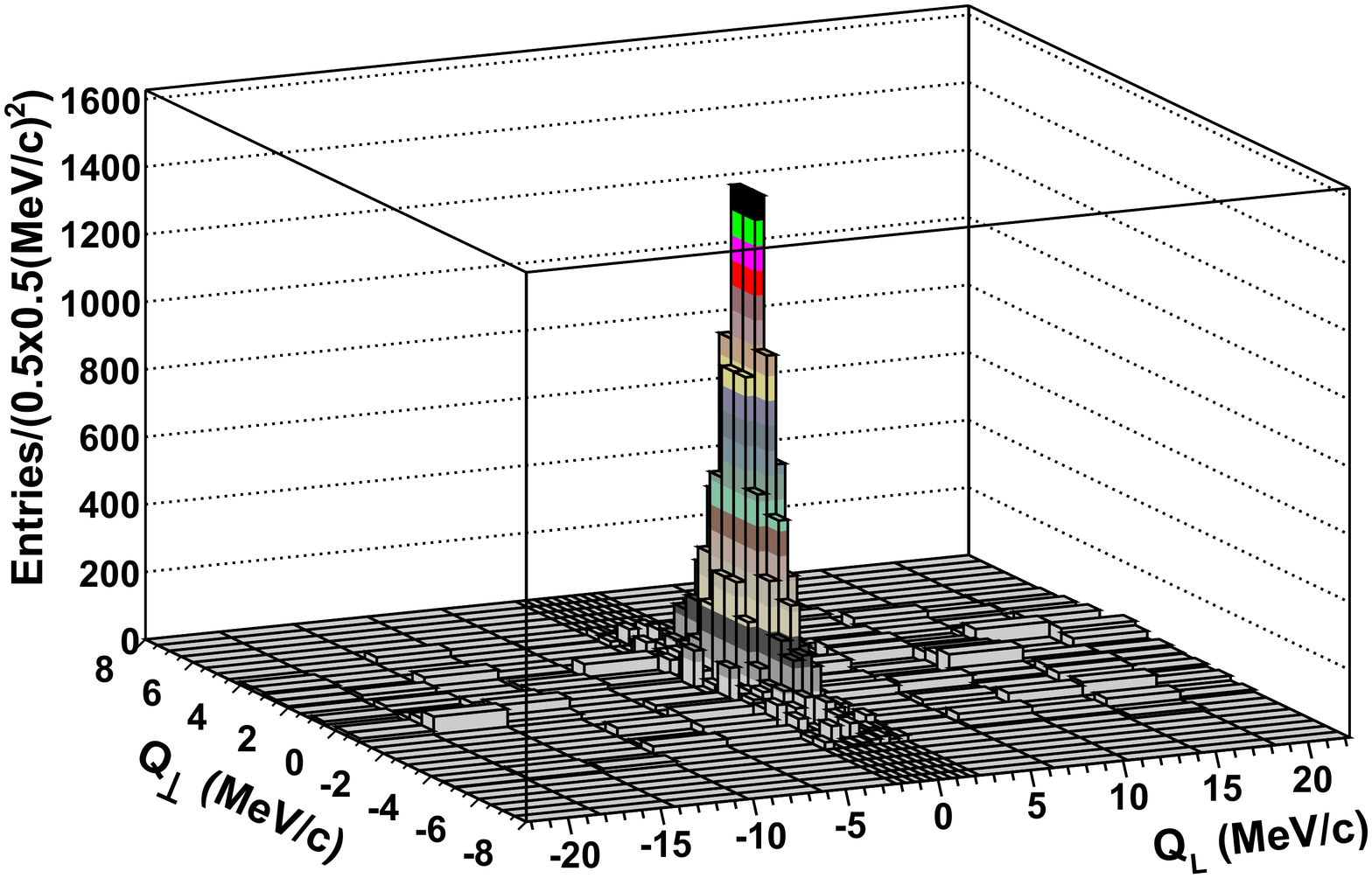}
\caption{Coulomb subtracted two-pion correlation function measured 
  in the ($Q_{\perp},Q_L$) plane, showing the pionium signal.
  $Q_{\perp}$ is the signed projection of $\vec{Q}$ into a generic
  transverse axis
%(arbitrary under the assumption of rotational symmetry)
  (azimuthal invariance is ensured by the absence of beam and target
  polarization) }
\label{qlego}
\end{figure} 

The only free parameters in~(\ref{chi2}) are the number of detected
atomic pairs ($n_{\mathrm{A}}^{\mathrm{rec}}$) and the fraction of
non-Coulomb/Coulomb pairs
($N_{\mathrm{nC}}^{\mathrm{rec}}/N_{\mathrm{C}}^{\mathrm{rec}}$). The
minimization is performed in two-dimensional space $|Q_L|<15$
MeV/$c$, $Q_T<5$ MeV/$c$, for values of the total pair momentum $p$
between 2.6 and 6.8 GeV/$c$~\cite{note0603}. A constraint on the total
number of reconstructed prompt pairs is applied such that
$N_{\mathrm{pr}}(1-\omega_{\mathrm{acc}})=N_{\mathrm{C}}^{\mathrm{rec}}+N_{\mathrm{nC}}^{\mathrm{rec}}+n_{\mathrm{A}}^{\mathrm{rec}}$.

In Figs.~\ref{qldata} and \ref{qtdata}, the $|Q_L|$ and $Q_T$
projections of the experimental prompt $\pi^+\pi^-$ spectrum are shown
in comparison to the fitted simulated background spectrum ($F_A$ = 0).
After subtraction of the $F_B$ background, the experimental $A_{2\pi}$
signal emerges at small values of $|Q_L|$ (Fig.~\ref{qldata}) and
$Q_T$ (Fig.~\ref{qtdata}) and can be compared with the simulated $F_A$
signal. As expected, multiple-scattering in the target and upstream
detectors broadens the $Q_T$ signal shape. This is clearly shown in
the 2-dimensional plot of Fig.~\ref{qlego}.  The overall agreement
between the best-fit experimental and simulated spectra is excellent,
over the entire $Q_T, Q_L$ domain.  
%% For the sake of consistency, the
%% global fit~(\ref{chi2}) has been independently repeated for seven
%% different 600 MeV/$c$ bins of the pionium momentum $p$. In all cases
%% the fit quality is again excellent.

\section{Pionium breakup probability}

The pionium breakup probability, $P_{\mathrm{br}}$, is defined as the
ratio $n_{\mathrm{A}}/N_{\mathrm{A}}$ between the number
$n_{\mathrm{A}}$ of observed pairs from pionium ionization caused by
target atoms and the total number $N_{\mathrm{A}}$ of pionium atoms
formed by final-state interaction.  The latter can be inferred by
quantum mechanics from the number of Coulomb-interacting pairs
measured at low $Q$ according to the expression~\cite{Nem85}
\begin{equation}\label{atoms}
\frac{N_{\mathrm{A}}(\Omega)} {N_{\mathrm{C}}(\Omega)} = 
\frac{(2\pi M_\pi \alpha)^3} {\pi}
\cdot \frac{\sum_{n=1}^{\infty} 1/n^3} {\int_{\Omega}
  A_{\mathrm{C}}(Q) d^3Q} = 
K^{\mathrm{th}}(\Omega),
\end{equation}
where $\Omega$ is the 
%kinematic 
domain of integration 
%\newline 
$|Q_L|<2$~MeV/$c$ and $Q_T<5$~MeV/$c$, yielding
$K^{\mathrm{th}}$~=~0.1301. Differences in detector acceptance and
reconstruction efficiency for $n_{\mathrm{A}}$ and $N_{\mathrm{C}}$
pairs, $\epsilon_{\mathrm{A}}$ and $\epsilon_{\mathrm{C}}$
respectively, are taken into account by correcting the theoretical
factor~$K^{\mathrm{th}}$ as
\begin{equation}\label{kfactor}
K^{\mathrm{exp}}(\Omega) = K^{\mathrm{th}}(\Omega) 
\frac {\epsilon_{\mathrm{A}}(\Omega)} {\epsilon_{\mathrm{C}}(\Omega)}.
\end{equation}
Those differences arise mainly from the lesser resolution of the
upstream detectors for identifying close tracks at very low $Q_T$.
This occurs more frequently for atomic pairs than for Coulomb pairs.

The breakup probability is thus determined as
\begin{equation}\label{atoms1}
P_{\mathrm{br}} = \frac{n_{\mathrm{A}}}{N_{\mathrm{A}}} = 
\frac{n_{\mathrm{A}}^{\mathrm{rec}}(\Omega)} 
{N_{\mathrm{C}}^{\mathrm{rec}}(\Omega)}  
\cdot \frac{1}{K^{\mathrm{exp}}(\Omega)}.
\end{equation}

The momentum-dependent $K^{\mathrm{exp}}$ factor~(\ref{kfactor}) has been
%was  
calculated from fully reconstructed Monte Carlo atomic and Coulomb pairs. 
\begin{figure}
\includegraphics[viewport=5mm 89mm 181mm 194mm,width=\columnwidth]{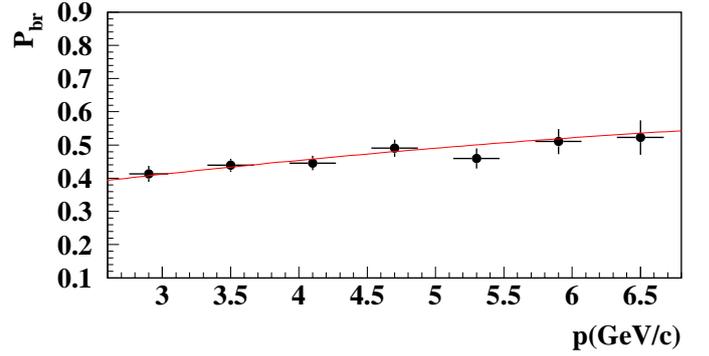}
\caption{The dependence of the measured $P_{\mathrm{br}}$, averaged 
  over all data sets, from the pionium laboratory momentum and the
  Monte Carlo prediction corresponding to the ground-state lifetime
  of $3.15\cdot10^{-15}$s obtained from the best fit.}
\label{dependence}
\end{figure}
Using~(\ref{kfactor}) and~(\ref{atoms1}), 35 independent
$P_{\mathrm{br}}$ values are obtained for the five independent data
sets and for seven 600~MeV/$c$ wide bins of the $A_{2\pi}$ momentum
from 2.6 to 6.8~GeV/$c$, by appropriately folding the momentum
dependence of $K^{\mathrm{exp}}$.

\begin{table*}[thb]
\caption{Fit results for $Q_T<5$MeV/$c$ and $|Q_L|<15$MeV/$c$.}
\label{table1}
\begin{tabular}{@{}|l|llllll|}
\hline
Ni, \hspace{0.3cm} $p_{\mathrm{beam}}$ \vphantom{$A^{A^A}$} &
$\chi^2/\mathrm{ndf}$ & $n_{\mathrm{A}}$ & 
$N_{\mathrm{C}}$ & $N_{\mathrm{nC}}$ & $N_{\mathrm{acc}}$ & 
$P_{\mathrm{br}}$ \\
\hline \vphantom{$A^{A^A}$} 
94$~\mu$m, 24~GeV/$c$ & 2127/2079 & 6020$\pm$216 & 
546003$\pm$4549 & 45624$\pm$4501 & 63212$\pm$208 & 0.441$\pm$0.018 \\
98$~\mu$m, 24~GeV/$c$ & 4288/4149 & 9321$\pm$274 & 
828554$\pm$5811 & 93148$\pm$5754 & 98499$\pm$255 & 0.452$\pm$0.015 \\
98$~\mu$m, 20~GeV/$c$ & 4257/4144 & 5886$\pm$210 & 
496820$\pm$4441 & 60867$\pm$4397 & 59392$\pm$144 & 0.472$\pm$0.020 \\
combined samples &   & 21227$\pm$407 & 
1871377$\pm$8613 & 199639$\pm$8526 & 221103$\pm$359 &  \\ 
\hline
\end{tabular}
\end{table*}

In Table~\ref{table1} the fitted yields are given for the different
momentum-averaged data sets. Overall, more than $2 \cdot 10^4$ atomic
pairs have been detected. The reported $P_{\mathrm{br}}$ values are
only indicative of the amount of variation expected with respect to
the different experimental conditions, and they are not used in the
final momentum-dependent fit.

A slight increase of the measured $P_{\mathrm{br}}$ with increasing
pionium momentum is observed in Fig.~\ref{dependence} (data points),
which is a consequence of the longer decay path, and hence the greater
breakup yield, expected at higher atom momenta.  The continuous curve
represents the predicted evolution of $P_{\mathrm{br}}$ with pionium
laboratory momentum, for the value of the pionium ground-state
lifetime $\tau = 3.15\cdot10^{-15}$~s obtained from this analysis.

\begin{figure}[htb]
\includegraphics[viewport=21mm 79mm 175mm 206mm,width=\columnwidth]{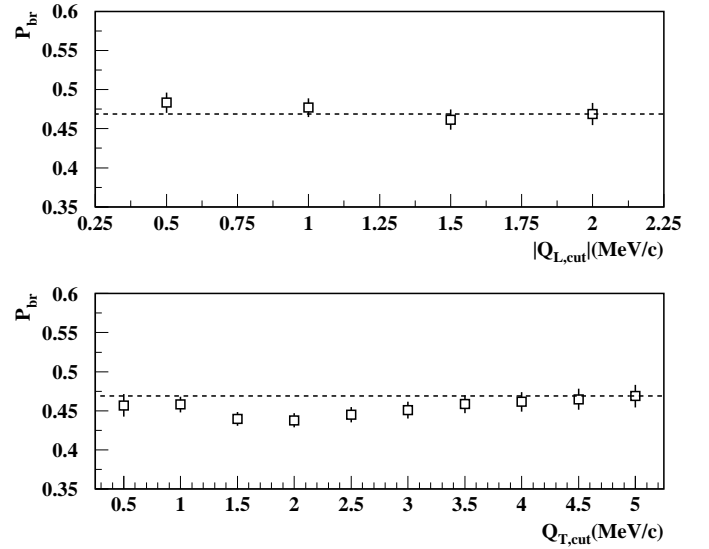}
\caption{Stability of the average $P_{\mathrm{br}}$ with respect to 
  variation of the: (top) $|Q_L|$ (for $Q_T <$ 5 MeV/$c$ ) and
  (bottom) $Q_T$ (for $|Q_L|<$~2 MeV/$c$) integration limits, in 0.5
  MeV/c bins.}
\label{stability}
\end{figure}

The dependence of the $A_{2\pi}$ breakup probability on the specific
choice of the integration domain $\Omega$ has been verified. The
measured $P_{\mathrm{br}}$, averaged over the data sets, is indeed
very stable versus variations of the $|Q_L|$, $Q_T$ integration limits
as shown in Fig.~\ref{stability}.

\section{Results and systematic errors}

A detailed assessment of the systematic errors affecting the
$P_{\mathrm{br}}$ measurement has been carried out, considering all
known sources of uncertainty in the simulation and in the theoretical
calculations.  The largest systematic error comes from a $\sim 1\%$
uncertainty in the multiple-scattering angle inside the Ni target
foil which induces a $\pm 0.0077$ error on $P_{\mathrm{br}}$.  The
momentum smearing correction can increase $P_{\mathrm{br}}$ by $\sim
2\%$ and thus produce a $\pm 0.0026$ systematic error.  The
double-track resolution at small angles can change $P_{\mathrm{br}}$
by $1.1\%$ and generate a systematic error of $\pm 0.0014$.  The
admixture of $K^+K^-$ changes $P_{\mathrm{br}}$ by $\sim1\%$. The
uncertainty on such contamination is $15\%$ and produces a systematic
error of $\pm 0.0011$ on $P_{\mathrm{br}}$.  The finite-size
correction to the point-like approximation creates a maximum $0.8\%$
variation of the simulated yield of Coulomb pairs and a systematic
error of $\pm 0.0011$ on $P_{\mathrm{br}}$. The influence of the
final-state strong interaction on the $\tau$ dependence of
$P_{\mathrm{br}}$ is negligible~\cite{Zhab08, Led08}.  The trigger
response efficiency was measured using minimum-bias events and
accidental pairs from calibration runs. The efficiency is high and
quite uniform in the selected $Q_T$, $Q_L$ domain and it drops by
$\sim 2\%$ per MeV/$c$ at ~$|Q_L|>15$ MeV/$c$. The simulated and
experimental trigger efficiencies agree to better than $0.5\%$, in the
same $|Q_L|$ range. This maximum deviation increases the breakup
probability by $\sim 3\%$ and thus produces a systematic error of $\pm
0.0004$.  Background hits in the upstream spectrometer region,
generated by beam and secondary interactions in the target region, are
the source of a $\pm 0.0001$ systematic error on $P_{\mathrm{br}}$.
The effect of the lower purity of the $94~\mu$m Ni target foil
compared to the $98~\mu$m is an underestimation of $P_{\mathrm{br}}$
by $\sim 1.1 \%$. This corresponds to a systematic error of $\pm
0.0013$ for the corresponding data set.

The dependence of $P_{\mathrm{br}}$ on the atom lifetime $\tau$, its
momentum, and the target parameters has been extensively studied for
several target materials, both by exactly solving the system of
transport equations~\cite{Afan96, Zhab08} describing the $A_{2\pi}$
excitation/de-excitation, breakup and annihilation, and by
simulating~\cite{Santa03} the $A_{2\pi}$ propagation in the target
foil.  The precision reached by these calculations is at the level of
1$\%$ \cite{Taras91}, which is reflected in a $\pm$0.0042 systematic
error on $P_{\mathrm{br}}$ for a lifetime $\tau=3.15\cdot10^{-15}$~s.
%
%% The result of these calculations defines two functions
%% $P_{\mathrm{br}}(\tau, p)$, one for each of the targets used.  The
%% functions $P_{\mathrm{br}}(\tau, p)$ are further convoluted with the
%% experimental momentum spectra of Coulomb pairs inside seven equal
%% ranges of the pionium laboratory momentum, from 2.6 to 6.8~GeV/$c$.
%
The result of these calculations defines three functions
$P_{\mathrm{br}}(\tau, p)$, one for each of the combinations of target
thickness and beam momentum.  The functions $P_{\mathrm{br}}(\tau, p)$
are further convoluted with the experimental momentum spectra of
Coulomb pairs inside the seven (600 MeV/c wide) momentum slices of the
pionium laboratory momentum, from 2.6 to 6.8~GeV/$c$. This approach
ensures that within each slice the non-linear dependence of
$P_{\mathrm{br}}(\tau)$ on the laboratory momentum is negligible.

Coulomb pairs, which have a momentum spectrum similar to that of
atomic pairs, are taken from prompt pairs in the $\vec{Q}$ region away
from the $A_{2\pi}$ signal, after subtraction of the non-Coulomb
contribution. The values of the systematic errors are summarized in
Table~\ref{table2}.

\begin{table}[htb]
%%\caption{Summary of systematic errors.}
\caption{Summary of systematic errors on $P_{\mathrm{br}}$.}
\label{table2}
\begin{center}
\begin{tabular}{|l|l|}
\hline
source & $\sigma$ \\
\hline
multiple scattering     & $\pm 0.0077$ \\
momentum smearing       & $\pm 0.0026$ \\
double-track resolution & $\pm 0.0014$ \\
$K^+K^-$ and $\mathrm{p\bar{p}}$ & $\pm 0.0011$ \\
trigger simulation & $\pm 0.0004$ \\
background hits & $\pm 0.0001$ \\
target impurity & $\pm 0.0013$ \\
finite size & $\pm 0.0011$ \\
calculation of $P_{\mathrm{br}}(\tau)$ & $\pm 0.0042$ \\
\hline
Overall error & $\pm 0.0094$\\
\hline
\end{tabular}
\end{center}
\end{table}

\section{Conclusions}

\begin{figure}[htb]
\includegraphics[width=\columnwidth]{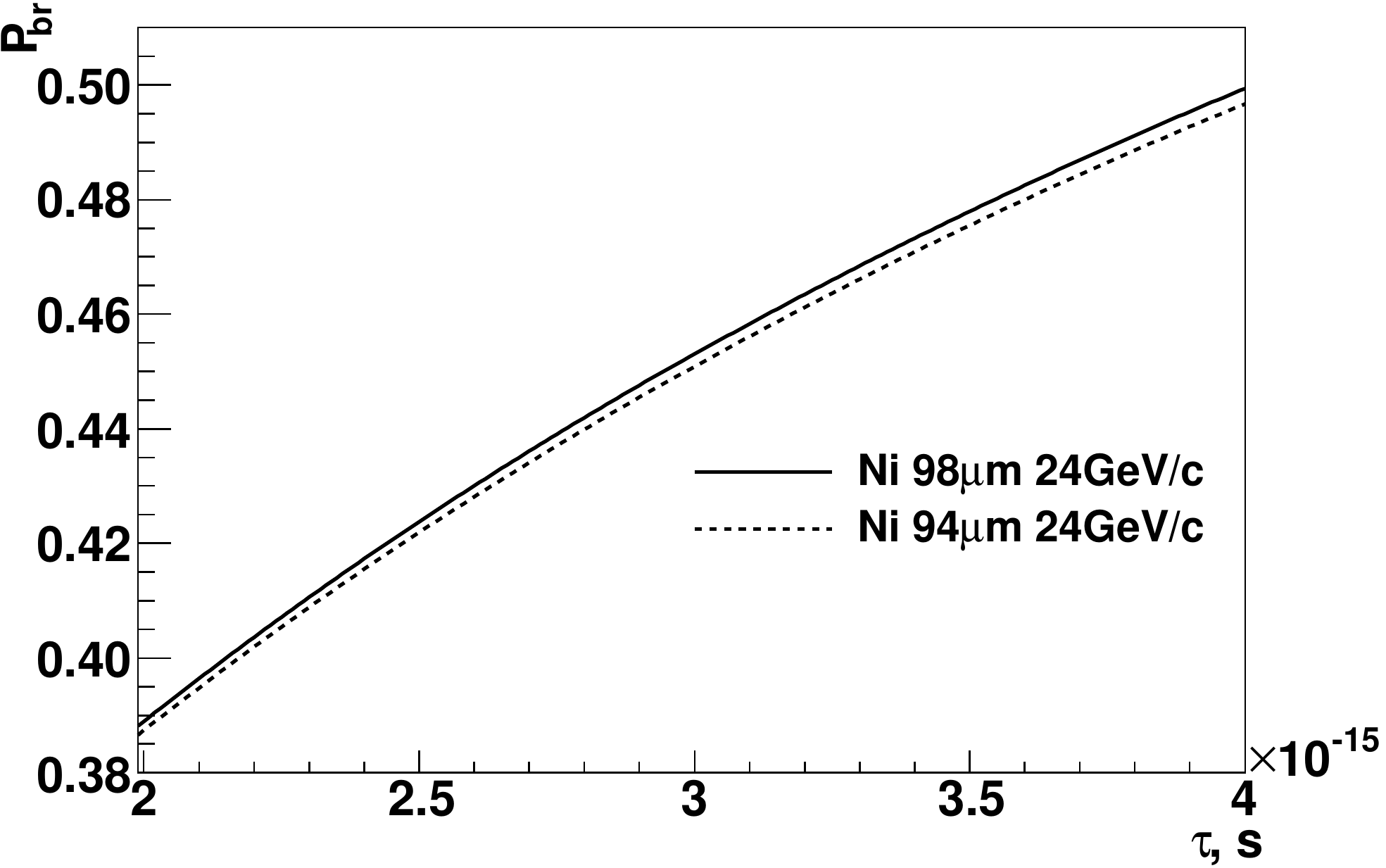}
\caption{Function $P_{\mathrm{br}}(\tau)$ corresponding to the
  dependence on pionium lifetime of the breakup probability for
  different targets.}
\label{pbr}
\end{figure}

Finally, the $P_{\mathrm{br}}$ measurements, obtained for the
different experimental conditions and $A_{2\pi}$ momentum ranges, and
their predicted $P_{\mathrm{br}}(\tau, p)$ values (see Fig.\ref{pbr}),
were used in a maximum likelihood fit of the lifetime
$\tau$~\cite{daniel08}.  Both statistical and systematic uncertainties
were taken into account in the maximization procedure.

Our final measurement of the ground-state $A_{2\pi}$ lifetime yields
$\tau = \left(\left.3.15_{-0.19}^{+0.20}\right|_\mathrm{stat}
  \left.{}_{-0.18}^{+0.20}\right|_\mathrm{syst}\right) \times
10^{-15}$~s.

Taking into account $A_{2\pi} \rightarrow \gamma\gamma$ and using
formula (\ref{eq:gasser}), we obtain the $\pi\pi$ scattering length
difference
\begin{equation}
  \label{eq:a0a2}
|a_0-a_2| =
\left(\left.0.2533^{+0.0080}_{-0.0078}\right|_\mathrm{stat}
  \left.{}^{+0.0078}_{-0.0073}\right|_\mathrm{syst}\right)
M_{\pi^+}^{-1},
\end{equation}
where the systematic error includes the 0.6\%
uncertainty induced by the theoretical uncertainty on the correction
$\delta$.

In conclusion, we have measured the ground-state lifetime of pionium
with a total uncertainty of $\sim9\%$. This represents the most
accurate lifetime measurement ever obtained and has allowed us to
determine the scattering length difference $|a_0-a_2|$ with a
$\sim4\%$ accuracy. Our result is in agreement with values of the
scattering lengths obtained from $K_{e4}$~\cite{Bat10} and $
K_{3\pi}$~\cite{Bat09} decay measurements using a completely different
experimental approach.

\section{Acknowledgments}

We are indebted to CERN for continuous support and the PS team for the
excellent performance of the accelerator. We acknowledge the computing
help from CESGA (Spain). This work was funded by CERN, INFN (Italy),
INCITE and MICINN (Spain), IFIN-HH (Romania), the Ministry of
Education and Science and RFBR grant 01-02-17756-a (Russia), the
Grant-in-Aid from JSPS and Sentanken-grant from Kyoto Sangyo
University (Japan).


\begin{thebibliography}{99}

\bibitem{Uretsky61} J.~Uretsky and J.~Palfrey, Phys. Rev. 121 (1961) 1798.

\bibitem{Gasser08} J.~Gasser et al., Phys. Rep. 456 (2008) 167
%; arXiv:0711.3522 [hep-ph]
  
\bibitem{Gasser01} J.~Gasser et al., Phys. Rev. D64 (2001) 016008
%; hep-ph/0103157.

\bibitem{Wein79} S.~Weinberg, Physica A96 (1979) 327.

\bibitem{Gasser85} J.~Gasser and H.~Leutwyler, Nucl. Phys. B250 (1985) 465.

\bibitem{Colan01NP} G.~Colangelo et al., Nucl. Phys. B603 (2001) 125.
 
\bibitem{Knecht95} M.~Knecht et al., Nucl. Phys. B457 (1995) 513.

\bibitem{Bat10} J.R. Batley et al., Eur. Phys. J. C70 (2010) 635.;\\
G. Colangelo, J. Gasser, A. Rusetsky, Eur. Phys. J. C59 (2009) 777.

\bibitem{Bat09} J.R. Batley et al., Eur. Phys. J. C64 (2009) 589.;\\
G. Colangelo et al. Phys. Lett. B638, 187 (2006). 

%\bibitem{Cabi04} N.~Cabibbo, Phys. Rev. Lett, 93 (2004) 121801. 
  
\bibitem{Nem85} L.L.~Nemenov, Sov. J. Nucl. Phys. 41 (1985) 629.

\bibitem{Adeva03} B.~Adeva et al., Nucl. Instrum. and Meth. A515 (2003) 467.

\bibitem{Afan94} L.G.~Afanasyev et al., Phys. Lett. B338 (1994) 478.

\bibitem{Adeva05} B. Adeva et al., Phys. Lett. B619 (2005) 50.

\bibitem{Afan96} L.G.~Afanasyev and A.V.~Tarasov, Phys. At. Nucl. 59 (1996) 2130.

\bibitem{Santa03} C.~Santamarina et al., J. Phys. B36 (2003) 4273.
  
\bibitem{Gorch96} O.E.~Gorchakov et al., Phys. At. Nucl. 59 (1996) 1942.
  
\bibitem{Zhab07} M.V.~Zhabitsky, DIRAC NOTES 2007-01, 2007-11;\\
http://cdsweb.cern.ch/record/1369660

\bibitem{Zhab08} M.V.~Zhabitsky, Phys. At. Nucl. 71 (2008) 1040
%; arXiv:0710.4416 [hep-ph] 

\bibitem{Afan99} L.~Afanasyev et al., J. Phys. G25
(1999) B7.  

\bibitem{Afan02} L.~Afanasyev et al., Phys. Rev. D 65 (2002) 096001
%; hep-ph/0109208.
  
\bibitem{Schum02} M.~Schumann et al., J. Phys. B35 (2002) 2683.
  
\bibitem{Led08} R.~Lednicky, J. Phys. G: Nucl. Part. Phys. 35 (2008) 125109.

\bibitem{Chl09} P.V.~Chliapnikov and V.M. Ronjin, J. Phys. G: Nucl. Part. Phys. 36
(2009) 105004.

\bibitem{note0702} B.~Adeva, et al., DIRAC NOTE 2007-02;\\
http://cdsweb.cern.ch/record/1369659

\bibitem{Lynch91} G.R. Lynch and O.I. Dahl, Nucl. Instrum. and Meth. B58, 6 (1991). 

\bibitem{note0806} A. Dudarev et al., DIRAC NOTE 2008-06;\\
http://cdsweb.cern.ch/record/1369639

\bibitem{Note-05-16i} B. Adeva et al., DIRAC NOTE 2005-16;\\
http://cdsweb.cern.ch/record/1369675

\bibitem{note0603} B.~Adeva et al., DIRAC NOTE 2006-03;\\
http://cdsweb.cern.ch/record/1369664
 
\bibitem{Taras91} A.V.~Tarasov and I.U.~Khristova, JINR-P2-91-10 (1991).

\bibitem{daniel08} D.~Drijard and M.~Zhabitsky, DIRAC NOTE 2008-07;\\
http://cdsweb.cern.ch/record/1369638

\end{thebibliography}
\end{document}